# Bending sound in graphene: origin and manifestation

V.M. Adamyan[1], V.N. Bondarev[1], V.V. Zavalniuk[1,2,*]

[1]Department of Theoretical Physics,
Odessa I.I. Mechnikov National University, 2 Dvoryanska St., Odessa 65026, Ukraine
[2]Department of Fundamental Sciences,
Odessa Military Academy, 10 Fontanska Road, Odessa 65009, Ukraine

**Abstract**

It is proved that the acoustic-type dispersion of bending mode in graphene is generated by the fluctuation interaction between in-plane and out-of-plane terms in the free energy arising with account of non-linear components in the graphene strain tensor. In doing so we use an original adiabatic approximation based on the alleged (confirmed *a posteriori*) significant difference of sound speeds for in-plane and bending modes. The explicit expression for the bending sound speed depending only on the graphene mass density, in-plane elastic constants and temperature is deduced as well as the characteristics of the microscopic corrugations of graphene. The obtained results are in good quantitative agreement with the data of real experiments and computer simulations.

## 1. Introduction

It is well-known that the lattice dynamics of "zero-thickness" crystals has a principal feature, which is not inherent to bulk solids. This is the logarithmic in a 2D lattice area growth of the mean-square atomic displacement at non-zero temperatures (the Peierls-Landau theorem [1]). A more "dangerous" consequence of low dimension, which might appear in 2D crystals, is connected with the classical "membrane" effect [2]: in a suspended (free-standing) state, the dispersion law of the so-called bending (out-of-plane) atomic vibrating mode $\omega_B = \sqrt{\kappa/\rho}\, q^2$ is quadratic upon the wave-number $q$; $\kappa > 0$ is the bending rigidity [2] and $\rho$ is the mass density of 2D crystal. Then the mentioned mean-square displacement found using this law may be proportional to the 2D crystal area [3] (see also [4]). Such "membrane" effect should first of all manifest itself in graphene [5] whose comprehensive study was stimulated by work [6].

Meanwhile, the first results of numerical simulation of the normal-normal correlation function for the graphene fluctuating surface [7] showed that for small ($q < 0.1$ Å$^{-1}$) wave numbers it does not diverge anymore or tends to a saturation (see also papers [8, 9] on simulations of the height-height correlation functions for graphene sheets). If so, then actually the eigenfrequencies of long-wave bending vibrations in graphene decrease as $q \to 0$ not faster than linearly in $q$ like those for the in-plane vibrations. Hence, the mean-square atomic displacement in a graphene sheet depends on the sheet area at most logarithmically. However, consequent calculations for large graphene system using a modified Monte Carlo method and molecular dynamics (MD) simulations [10] did not show any saturation of the normal-normal correlation function at least for $q > 0.02$ Å$^{-1}$ (see also [11]) that could indicate a lack of low-frequency sound segment in the graphene bending mode.

Nevertheless, the linear dispersion of the bending mode in graphene at $q \to 0$ was established in the recent papers [12, 13]. In [12] this was done within the quantum theory of crystalline membrane with account of cubic interactions between in-plane and out-of-plane displacement fields and a quartic local interaction for the out-of-plane displacements. In [13] starting from a discrete atomistic model of a monolayer crystal with anharmonic coupling of third and fourth orders, a dependence $\omega_B = s_B q$ has been also obtained. It is worth mentioning the work [14], in which the linear dispersion of the bending mode at $q \to 0$ is a result of coupling between structural and electronic degrees of freedom in graphene. The found in [14] $s_B \approx 1$ km/s at 300 K turned out 15–20 times less than the in-plane sound speeds in graphene. Note that close estimate $s_B \approx 1.6$ km/s has

---

been obtained in [15] by analogy between the phonon dispersion curves in graphene and experimental results for graphite.

It is worth mentioning the recent work [16], in which the linear dispersion of the out-of-plane acoustic mode in graphene was obtained by means of classical MD simulations. According to the results of [16], the bending sound speed $s_B$ demonstrates very small size effect and changes from 0.4 km/s at 300 K to 0.6 km/s at 2000 K.

Besides, there are experimental facts that give grounds to suggest that $s_B \neq 0$ in graphene. To give a consistent explanation of the temperature dependence of the electron mobility in graphene, it was suggested in [17] that the flexural (out-of-plane) phonons are a major source of electron scattering in suspended graphene. At the same time to match the experimental data the authors of [17] suggested the existence of some (in fact, "frozen") in-plane strain in graphene (see also [18], where the ripples or microscopic corrugations of a graphene sheet are discussed). The presence of these static strains results in the linear dispersion of the bending mode at $q \to 0$. It is worth mentioning that the existence of structural corrugations ("intrinsic microscopic roughening" [19]) of the free-standing graphene with an amplitude ~ 1 Å and a characteristic wave-length ≈ 50 Å had been observed in the transmission electron microscopy experiments [19–21].

Thus, neither the experimental data nor results of numerical simulation of the structure and phonon spectra of free standing graphene sheet demonstrate any indications of "membrane" effect in the graphene out-of-plane vibrations at $q \to 0$. So, the convincing theoretical arguments in favor of the sound-like long-wave dispersion for the bending vibrations of graphene-like 2D crystals are needed.

Strikingly small value of $s_B$ in comparison with in-plane sound speeds of graphene indicates that the origin of the bending sound differs radically from that of in-plane modes. In the present paper, using transparent physical arguments we show that the long-wave region of the bending mode spectrum must necessarily have linear in wave number dispersion. This result is obtained through the account of the terms represented by products of bilinear combinations of both in-plane and out-of-plane deformations in the graphene elastic free energy. Note that such terms are usually ignored, and their accounting is a key point of our approach to the theory of elastic properties of quasi-2D solids. Using the known values of elastic and structure parameters of graphene, we derive the bending sound speed $s_B$ for arbitrary temperatures without introducing any additional fitting parameters. What matters, the derived formula for $s_B$ is independent of the graphene sample size and is expressed only through its in-plane moduli (and also third-order elastic constants) and temperature. Note that combining results of our approach with result of [12] one can verify that the considered in [12] cubic and quartic terms, which we ignored in the graphene free energy when deriving the expression for $s_B$, do not change the found expression at least at high temperatures. The corresponding analysis permits also to renormalize the bending rigidity coefficient $\kappa$.

The approach developed in the present work allows to reproduce with reasonable accuracy the main results of [16] referencing only to one value of $s_B$ at a certain temperature. Besides, the mean-square out-of-plane displacement for free standing graphene of given linear size is obtained and fluctuation corrugations of graphene are also described. The theory demonstrates quantitative agreement with the experimental data and the results of computer simulations in wide temperature interval. In principle, the results of the paper may also be used for study of the dynamics of graphene-like crystals: silicene, germanene, graphane etc. (see [22–24]).

## 2. The "sound" segment of the free-standing graphene bending mode

For study of long-wave mechanical vibrations in graphene we use a continuum model of elastic 2D plane in 3D space. Let $\mathbf{r} = (x, y)$ be the radius-vectors of graphene points in equilibrium, $\mathbf{u}(\mathbf{r})$ and $w(\mathbf{r})$ are corresponding in-plane and out-of-plane components of displacement vectors, respectively; $\dot{\mathbf{u}}(\mathbf{r})$, $\dot{w}(\mathbf{r})$ are time-derivatives of $\mathbf{u}(\mathbf{r})$, $w(\mathbf{r})$.

Thereafter the "Hamiltonian" for long-wave mechanical vibration in hexagonal graphene can be written in the form:

$$\boldsymbol{H} = \int d\mathbf{r} \left\{ \frac{\rho}{2} \left[ \dot{\mathbf{u}}^2(\mathbf{r}) + \dot{w}^2(\mathbf{r}) \right] + \frac{\lambda}{2} \varepsilon_{\alpha\alpha}(\mathbf{r}) \varepsilon_{\beta\beta}(\mathbf{r}) + \mu \varepsilon_{\alpha\beta}(\mathbf{r}) \varepsilon_{\alpha\beta}(\mathbf{r}) + \frac{\kappa}{2} \left[ \nabla^2 w(\mathbf{r}) \right]^2 \right\}, \qquad (1)$$

where

$$\varepsilon_{\alpha\beta}(\mathbf{r}) = \frac{1}{2}\left[\frac{\partial u_\alpha(\mathbf{r})}{\partial r_\beta} + \frac{\partial u_\beta(\mathbf{r})}{\partial r_\alpha} + \frac{\partial u_\gamma(\mathbf{r})}{\partial r_\alpha}\frac{\partial u_\gamma(\mathbf{r})}{\partial r_\beta} + \frac{\partial w(\mathbf{r})}{\partial r_\alpha}\frac{\partial w(\mathbf{r})}{\partial r_\beta}\right] \quad (2)$$

are in-plane components of strain tensor [2] with $\alpha, \beta, \gamma$ running $x, y$ (summation in repeating subscripts is implied); $\nabla$ is the 2D gradient, $\rho$ is the 2D mass density, $\lambda > -\mu$ and $\mu > 0$ are Lamé coefficients. In (1) we also included the term with the bending rigidity $\kappa > 0$ which is usually taken into account when considering the flexural effects in membranes [2]. Note that this term for graphene as one-atom-thick 2D layer can not be directly considered in the framework of the elasticity theory for macroscopically "thin" plates [2]. Really, the formal expression for $\kappa$ in the classical elasticity theory contains the cube of a plate thickness [2] and in application to 2D graphene sheet of "zero" thickness a macroscopic interpretation of the parameter $\kappa$ becomes problematic. Nevertheless, graphene as quantum 2D lattice of carbon atoms with strong covalent bonds must possess a finite flexural rigidity due to a change of electron hybridization at microscopic bending of graphene [25]. In addition, a certain contribution to $\kappa$ is attributable to non-linear terms in (2) (below we obtain explicit expression for this contribution using the approach developed in [12]). However, the modeling of bending rigidity for multilayer graphene by formulas of the classical theory of elasticity may be justified [26].

Contrary to many papers on the topic we keep the *quadratic* terms $(\partial u_\gamma(\mathbf{r})/\partial r_\alpha)(\partial u_\gamma(\mathbf{r})/\partial r_\beta)$ in the expression (2) for the strain tensor, which are usually treated as small. This is the key point of our approach.

In the free standing graphene not affected by the action of external forces the strains can have the only oscillation nature (we, surely, disregard the boundary effects). So, in the first order of perturbation theory, the terms of kind

$$\left[\frac{\partial u_\alpha(\mathbf{r})}{\partial r_\beta} + \frac{\partial u_\beta(\mathbf{r})}{\partial r_\alpha}\right]\frac{\partial w(\mathbf{r})}{\partial r_\alpha}\frac{\partial w(\mathbf{r})}{\partial r_\beta}, \quad (3)$$

in (1), in fact, will give zero contribution into the free-energy of the long-wave out-of-plane deformations. Indeed, the linear in the in-plane phonons factor in (3) is "rapidly fluctuating" in comparison with the quadratic one related to phonons of the bending branch (cf. the discussion concerning the speeds of the corresponding modes in the Introduction; besides, formally by the dispersion law $\omega_B = \sqrt{\kappa/\rho}\, q^2$ at $q \to 0$ the speed $d\omega_B/dq \to 0$). Thus, during the period of the "fast" in-plane oscillations the factor $[\partial w(\mathbf{r})/\partial r_\alpha][\partial w(\mathbf{r})/\partial r_\beta]$ in (3) can be considered as constant, and then the average of the linear on the in-plane phonons factor is obviously zero. However, in the second order of perturbation theory the terms (3) give non-zero contribution into the free-energy of the out-of-plane mode (see below in this section).[1] Moreover, we will make sure below by a direct calculation that the resulting speed of bending sound mode $s_B \sim 1$ km/s is actually 15-20 times less than the speeds of the in-plane acoustic phonons. Just this fact allows us to develop some adiabatic approximation by separation of the acoustic vibrations of graphene into the "fast" (in-plane) and "slow" (out-of-plane) ones.

As regards the terms

$$\frac{\partial u_\gamma(\mathbf{r})}{\partial r_\alpha}\frac{\partial u_\gamma(\mathbf{r})}{\partial r_\beta}\frac{\partial w(\mathbf{r})}{\partial r_\alpha}\frac{\partial w(\mathbf{r})}{\partial r_\beta}, \quad (4)$$

also appearing in (1), their average over the "rapidly fluctuating" in-plane variable (designated further by the angle brackets) is non-zero at any temperature actually in the first order of perturbation theory. From (4) by symmetry we have

$$\left\langle \frac{\partial u_\gamma(\mathbf{r})}{\partial r_\alpha} \frac{\partial u_\gamma(\mathbf{r})}{\partial r_\beta} \right\rangle \frac{\partial w(\mathbf{r})}{\partial r_\alpha} \frac{\partial w(\mathbf{r})}{\partial r_\beta} = \frac{1}{2} \left\langle \frac{\partial u_\gamma(\mathbf{r})}{\partial r_\alpha} \frac{\partial u_\gamma(\mathbf{r})}{\partial r_\alpha} \right\rangle [\nabla w(\mathbf{r})]^2. \qquad (5)$$

---

[1] Note, that the above arguments suggest that the free-standing graphene does not undergo an external (for example, from the substrate) stress. The latter can create in graphene the static ("frozen") strains, due to which the terms like (3) "work" already in the first order of perturbation theory, that, in fact, was supposed in [17, 18].

Change for simplicity the real spectrum of the in-plane oscillation modes $\omega_{L,T}(q) = s_{L,T} q$ with the longitudinal $s_L = \sqrt{(\lambda+2\mu)/\rho}$ and the transversal $s_T = \sqrt{\mu/\rho}$ sound speeds by two modes $\omega_{L,T}(k) = s_\| q$ with the "average" speed of sound defined by the relation $2 s_\|^{-2} = s_L^{-2} + s_T^{-2}$ [1]. Then, as usually [1, 27], we pass to the 2D Fourier-representation for the displacements in the $xy$ plane: $\mathbf{u}(\mathbf{r}) = \sum_\mathbf{k} \mathbf{u}_\mathbf{k} \exp(i\mathbf{k}\mathbf{r})$ (with $\mathbf{u}_{-\mathbf{k}} = \mathbf{u}_\mathbf{k}^*$) and use the Debye model introducing the maximum wave-number $k_{max} = \sqrt{4\pi\rho/m}$, the maximum frequency $\omega_{max} \equiv s_\| k_{max}$ and the corresponding (connected with the in-plane phonons) Debye temperature $\theta_\| \equiv \hbar \omega_{max}$. Applying the standard technique of the solid state theory, i.e. taking the mentioned averages with account of Planck distribution function of in-plane phonons, we obtain (see, for example [27])

$$\left\langle \frac{\partial u_\gamma(\mathbf{r})}{\partial r_\alpha} \frac{\partial u_\gamma(\mathbf{r})}{\partial r_\alpha} \right\rangle = \frac{\hbar}{\pi \rho s_\|^4} \int_0^{\omega_{max}} \omega^2 \left( \frac{1}{e^{\hbar\omega/T}-1} + \frac{1}{2} \right) d\omega = \frac{\theta_\|^3}{6\pi\rho\hbar^2 s_\|^4} \left[ 1 + 6 \left( \frac{T}{\theta_\|} \right)^3 \int_0^{\theta_\|/T} \frac{\xi^2 d\xi}{e^\xi - 1} \right]. \qquad (6)$$

Now, collecting together all the averaged over the in-plane phonons terms of kind (5) with account of (6), we obtain from (1) the effective "Hamiltonian" for the linear elastic oscillations in graphene:

$$H\{\mathbf{u}, w\} = \int d\mathbf{r} \left\{ \frac{\rho}{2} [\dot{\mathbf{u}}^2(\mathbf{r}) + \dot{w}^2(\mathbf{r})] + \right.$$
$$\left. \frac{\lambda}{2} u_{\alpha\alpha}(\mathbf{r}) u_{\beta\beta}(\mathbf{r}) + \mu u_{\alpha\beta}(\mathbf{r}) u_{\alpha\beta}(\mathbf{r}) + \frac{B}{2} [\nabla w(\mathbf{r})]^2 + \frac{\kappa}{2} [\nabla^2 w(\mathbf{r})]^2 \right\}, \qquad (7)$$

where $u_{\alpha\beta}(\mathbf{r}) \equiv [\partial u_\alpha(\mathbf{r})/\partial r_\beta + \partial u_\beta(\mathbf{r})/\partial r_\alpha]/2$ and

$$B \equiv \frac{\lambda+\mu}{2} \left\langle \frac{\partial u_\gamma(\mathbf{r})}{\partial r_\alpha} \frac{\partial u_\gamma(\mathbf{r})}{\partial r_\alpha} \right\rangle = \frac{2(\lambda+\mu)\hbar\sqrt{\pi\rho}}{3 m^{3/2} s_\|} \left[ 1 + 6 \left( \frac{T}{\theta_\|} \right)^3 \int_0^{3\theta_\|/T} \frac{\xi^2 d\xi}{e^\xi - 1} \right]. \qquad (8)$$

Due to the presence of the *new* term with the bending "elasticity modulus" $B > 0$ in (7), the dispersion equation for the resulting out-of-plane bending mode has the form:

$$\omega_B^2(q) = s_B^2 q^2 + \frac{\kappa}{\rho} q^4, \qquad (9)$$

i.e. now it will contain the sound segment with the speed of the bending sound

$$s_B = \sqrt{B/\rho}. \qquad (10)$$

Pay attention to that the speed of the bending sound $s_B$ in our approach do not depend of the graphene sheet area (cf. the results of MD simulations [16]) and is expressed through the in-plane Lamé coefficients only. As far as we know, the explicit form (8) for the bending "elasticity modulus" $B$ has never been applied to 2D crystals.

Note that unlike $\lambda$ and $\mu$ the bending "elasticity modulus" $B$ has purely fluctuation origin. With increasing temperature $B$ also increases. By (8)

$$B = \begin{cases} \dfrac{2(\lambda+\mu)\hbar\sqrt{\pi\rho}}{3 m^{3/2} s_\|} \left[ 1 + 12\varsigma(3) \left( \dfrac{T}{\theta_\|} \right)^3 \right], & T \ll \theta_\|; \\[2ex] \dfrac{(\lambda+\mu)T}{m s_\|^2} \left[ 1 + \dfrac{1}{24} \left( \dfrac{\theta_\|}{T} \right)^2 + \ldots \right], & T \gg \theta_\|; \end{cases} \qquad (11)$$

where the Riemann zeta-function $\varsigma(3) = 1.202$. Thus, even at $T=0$ the elastic modulus $B$ (together with the speed $s_B$ of the bending sound) is finite due to the quantum zero oscillations of the in-plane modes. Note, that the asymptotic expressions in (11) actually are true if $T << \theta_\parallel/4$ or $T >> \theta_\parallel/4$ by the similar arguments as for 3D Debye model [1].

In the above derivation of equations (8), (9) we ignored in (1) the non-linear terms (3) and

$$\frac{\partial w(\mathbf{r})}{\partial r_\alpha} \frac{\partial w(\mathbf{r})}{\partial r_\alpha} \frac{\partial w(\mathbf{r})}{\partial r_\beta} \frac{\partial w(\mathbf{r})}{\partial r_\beta}. \tag{12}$$

Notice that such quartic constructions, as well as the terms of the form (3), previously were used to perform the renormalization of constants entering into "Hamiltonian" (1) (see, for example, [3, 4, 28, 29]). To see how their account would amend the function $\omega_B(q)$ let us combine (9) and the results of [12] on the dispersion equation for the out-of-plane mode. Note that in [12] contrary to the present work the terms (3) and (12) in "Hamiltonian" (1) were held but the terms (4) were discarded. To draw together the both contributions one can take the "Hamiltonian" (7) as unperturbed and add to it the combination of terms (3) and (12) as perturbation. Then the equation defining the resultant spectrum of the bending mode can be written in the form:

$$\omega_B^2(q) = \frac{B}{\rho} q^2 + \frac{\kappa}{\rho} q^4 + \frac{1}{\rho} \Sigma_q(T; B, \kappa; q),$$

where $\Sigma_q(T; B, \kappa; q)$ is the first-order self-energy term.

Applying similar arguments as in [12] yields for $T = 0$ the following expression for the first-order self-energy of the out-of-plane mode:

$$\Sigma_q(T=0; B, \kappa; q) = \frac{\hbar B}{8\pi \kappa^{3/2} \rho^{1/2}} \int_0^{k_{max}\sqrt{\kappa/B}} \left[ \frac{\lambda + 2\mu}{\sqrt{1+s^2} + \sqrt{(\lambda+2\mu)/B}} + \frac{\mu}{\sqrt{1+s^2} + \sqrt{\mu/B}} \right] s^2 ds \cdot q^2 +$$

$$\frac{3\hbar}{4\pi\sqrt{\kappa\rho}} \frac{\mu(\lambda+\mu)}{\lambda+2\mu} \ln\left[\sqrt{\frac{\kappa k_{max}^2}{B}} + \sqrt{\frac{\kappa k_{max}^2}{B}+1}\right] \cdot q^4 + \Delta(T=0; B, \kappa; q) \cdot q^4, \tag{13}$$

$$\Delta(T=0; B, \kappa; q) \le \frac{3\hbar k_{max}}{16\pi} \left\{ \frac{1}{\sqrt{(\lambda+2\mu)\rho}} \left[ \frac{1}{2} \cdot \frac{\lambda^2}{\lambda+2\mu} + 2(\lambda+\mu) \right] + \sqrt{\frac{\mu}{\rho}} \right\}.$$

The coefficient at $q^2$ in (13) is nothing but an addition $\Delta B$ to the bending "elasticity modulus" due to the anharmonic terms (3) and (12), so the resulting bending "elasticity modulus" is the sum $B+\Delta B$. Note that $\Delta B$ for $B << \mu$ just coincide at $T = 0$ with $B$ up to the inessential difference of $s_\parallel$ and the harmonic mean of longitudinal and transversal in-plane sound speeds.

Another consequence of (13) is that the coefficients at $q^4$ may be formally identified with the contributions (due to nonlinear terms in (2)) into the bending rigidity of graphene at $T = 0$, $\kappa_{nl}(T=0)$. Substituting $k_{max} = \sqrt{4\pi\rho/m}$ into this $\kappa_{nl}(T=0)$ and omitting $\Delta(T=0; B,\kappa; q)$ as negligible in comparison with another coefficient at $q^4$ we get

$$\kappa_{nl}(T=0) = \frac{3\hbar}{4\pi\sqrt{\kappa\rho}} \frac{\mu(\lambda+\mu)}{\lambda+2\mu} \ln\left[\sqrt{\frac{4\pi\kappa\rho}{Bm}} + \sqrt{\frac{4\pi\kappa\rho}{Bm}+1}\right]. \tag{14}$$

At finite temperatures ("$T\to\infty$", in terms of Ref. 12) following [12] one can find the first-order self-energy in the limit $q\to 0$:

$$\Sigma_q(T; B, \kappa; q \to 0) = \frac{3\mu(\lambda+\mu)T}{8\pi(\lambda+2\mu)B} \left[ \ln\left(\frac{B}{\kappa q^2}+1\right) + \frac{1}{2} \right] \cdot q^4. \tag{15}$$

From here it is clear that there is no renormalization of the bending sound speed (i.e., $B$) at rather high temperatures. Actually, at $q \to 0$, Eq. (15) gives the correction (due to nonlinear terms in (2)) to the bending rigidity of graphene:

$$\kappa_{nl}(T) \equiv \frac{1}{q^4}\Sigma_q(T; B, \kappa; q \to 0) = \frac{3\mu(\lambda+\mu)T}{8\pi(\lambda+2\mu)B}\left[\ln\left(\frac{B}{\kappa q^2}\right) + \frac{1}{2}\right]. \quad (16)$$

Taking into account (11), one can see that the dependence of $\kappa_{nl}(T)$ at high temperatures is only logarithmic (for constant $\lambda, \mu, \kappa$).

Now, let us make some numerical estimates, accepting for graphene at room temperature [17]: $\rho \approx 7.6 \times 10^{-8}$ g/cm$^2$, $\mu \approx 3\lambda \approx 9$ eV/Å$^2$, $m = 2 \times 10^{-23}$ g ($^{12}$C atomic mass), whence $s_L \approx 21$ km/s, $s_T \approx 14$ km/s, and the Debye temperature $\theta_\| \approx 2700$ K. In fact, values of $\rho$, $\lambda$ and $\mu$ depend on temperature and we took these dependencies from computer simulations [30]. Comparing expressions (14) and (16) it is easy to verify that the correction to the bending rigidity at high temperatures is definitely higher than that at $T = 0$, what correlates qualitatively with the temperature growth of the bending rigidity reported in [7, 16, 31–33].

It is important to note that the low- and high-temperature expressions (14) and (16), respectively, are in no way related to a vague "thickness" of 2D graphene sheet (concerning this "thickness" see, for example [5]).

Operating with the "Hamiltonian" (7) we took into account, as usually (see, for example, [17, 18]), non-linear terms in the strain tensor (2). Thus we have gone beyond the linear theory of elasticity, retaining, however, the formal structure of latter with only two Lamé coefficients $\lambda$ and $\mu$. Meanwhile, accounting of the terms like (4) demands, in general, also to include into elastic energy of graphene the all symmetry permissible anharmonic constructions of corresponding order. It is easy to see that the contributions of the form (4) are definitely contained in anharmonic constructions of the third (but not the fourth!) order in $\varepsilon_{\alpha\beta}(\mathbf{r})$ in the "Hamiltonian". In application to graphene, such third order constructions were written out in [34]. According to the experimental data [35], an effective value of the third order elastic modulus is negative. This actually means that in (8), instead of $\lambda + \mu$, should stand the expression $\lambda + \mu + C_3$, where $C_3 < 0$ is the contribution of the mentioned third order elastic terms into the effective bending modulus $B_{ef}$:

$$B_{ef} = \frac{2(\lambda+\mu+C_3)\hbar\sqrt{\pi\rho}}{3m^{3/2}s_\|}\left[1 + 6\left(\frac{T}{\theta_\|}\right)^3 \int_0^{\theta_\|/T} \frac{\xi^2 d\xi}{e^\xi - 1}\right], \quad (17)$$

which gives the bending sound speed through (10). In what follows we use for the bending mode the dispersion equation with $B_{ef}$ instead of $B$ in (13) and (15):

$$\omega_B^2(q) = \frac{B_{ef}}{\rho}q^2 + \frac{\kappa}{\rho}q^4 + \frac{1}{\rho}\Sigma_q(T; B_{ef}, \kappa; q). \quad (18)$$

Reasonable estimate for $C_3$ can be obtained by the analysis of results of [16] in which by classical MD simulations of the out-of-plane acoustic mode in graphene the growth of $s_B$ from 0.4 km/s at 300 K to 0.6 km/s at 2000 K was found. Using the room temperature value of $s_B$ [16] and of $\rho$, $\lambda$ and $\mu$ given above we find $C_3 \approx -8.7$ eV/Å$^2$.[2] Using for $C_3$ the same temperature dependence as that for $\rho$ and $\mu$ (see above), we obtain $s_B \approx 0.6$ km/s at 2000 K. Below we will see that the account of $C_3$ leads to quantitative agreement of our results and the data of numerical "experiments" on the out-of-plane fluctuation displacements (ripples) of free-standing graphene.

In Fig. 1, the solid line shows the temperature dependence of $s_B$ calculated by the formula $s_B = \sqrt{B_{ef}/\rho}$ with account of temperature dependencies of graphene density and elastic parameters. Besides, in Fig. 1 by the dashed line we represent the theoretical dependence of $s_B(T)$ calculated from Eqs. (8), (10). Despite the found temperature growth of $s_B$, it remains small in comparison with $s_L$ and $s_T$ even at high temperatures up to the melting point (about ~4500K [36]). This justifies the used adiabatic approach to the calculation of $B$.

---

[2] Unfortunately, the estimation of $C_3$ from [34, 35] is much higher in the absolute value than that above. So we use just the value $C_3 \approx -8.7$ eV/Å$^2$ in further calculations.

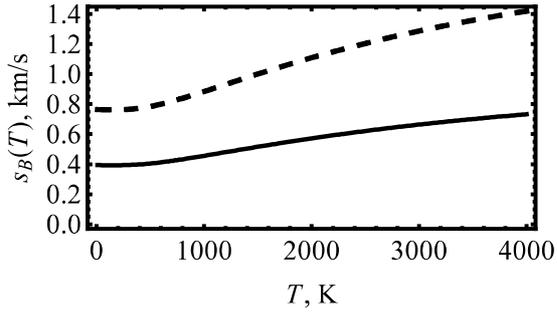

Fig. 1. Theoretical temperature dependences of bending sound speed $s_B$ without (dashed line) and with (solid line) account of third-order elasticity terms in the "Hamiltonian".

## 3. Manifestation of the bending mode in graphene thermodynamics

The effective "Hamiltonian" (7) and dispersion law (9) following from it allow us to calculate various thermal averages upon the distribution of the fluctuations of the out-of-plane displacements in graphene. Particularly,

$$\langle [\nabla w(\mathbf{r})]^2 \rangle = \frac{\hbar}{2\pi\rho} \int_0^{k_{max}} \frac{k^3 dk}{\omega_B(k)} \left[ \frac{1}{e^{\hbar\omega_B(k)/T} - 1} + \frac{1}{2} \right], \qquad (19)$$

and (18) as the dispersion equation is used here and below.

The calculation of the mean-square ripple "heights"

$$\langle w^2(\mathbf{r}) \rangle = \frac{\hbar}{2\pi\rho} \int_{k_{min}}^{k_{max}} \frac{k dk}{\omega_B(k)} \left[ \frac{1}{e^{\hbar\omega_B(k)/T} - 1} + \frac{1}{2} \right] \qquad (20)$$

demands introducing of a cut-off parameter $k_{min} > 0$ because of the divergence of integral in (20) at $T > 0$. To demonstrate this fact, let us calculate the integral in (20) neglecting the self-energy term in the dispersion equation (18). Considering $D$ as the diameter of a quasi-macroscopic sheet we take $k_{min} = \pi/D$. Then with this $k_{min}$, at "high" temperatures we get

$$\langle w^2(\mathbf{r}) \rangle = \frac{T}{4\pi B_{ef}} \ln\left[ \frac{1 + B_{ef} D^2 /(\pi^2 \kappa)}{1 + B_{ef} m /(4\pi\kappa\rho)} \right].$$

Although at "high" temperatures $\langle w^2(\mathbf{r}) \rangle$ diverges with increasing of $D$, this divergence is only logarithmic in contrast to the "membrane" case with the vanishing $B_{ef}$.[3] This logarithmic divergence of $\langle w^2(\mathbf{r}) \rangle$ is similar to that of $\langle u^2(\mathbf{r}) \rangle$ [1] (see discussion in Introduction). It is worth mentioning that this behavior agrees with the results of MD simulations of the out-of-plane graphene lattice dynamics [16].

Unfortunately, at present it is impossible to give a detailed comparison between the obtained theoretical results and real experimental data in view of the very limited information available in the literature concerning the dependences of the intrinsic corrugations peculiarities on the thermodynamic state of graphene. In this situation, numerical experiments are of the primary importance. However, before making a comparison of the temperature dependencies of $\sqrt{\langle w^2(\mathbf{r}) \rangle}$ calculated by (17) with those obtained by numerical simulation for graphene samples of different sizes one must note that usually such simulation is carried out for square pieces of graphene with typical lengths of the sides $L$ from 12.5 Å and higher (see, for example, [7, 30]).[4] As the simulations [37, 38] show, the Young's modulus $Y = \mu(3\lambda+2\mu)/(\lambda+\mu)$ [5] of a graphene sheet is $L$-dependent: the effective value of room temperature $Y$ increases about 1.6 times when increasing $L$ from 10 Å to 32 Å and then almost does not change with $L$. Besides, according to [5] $Y$ falls from 20 eV/Å$^2$ to 16 eV/Å$^2$ as temperature increases from room to melting.

---

[3] In the formal limit $B_{ef} \to 0$ Eq. (20) gives $\langle w^2(\mathbf{r}) \rangle = TD^2/(4\pi^3\kappa)$ (cf. [3]), what leads to a catastrophic growth of the out-of-plane fluctuations with increasing of the 2D crystal area. More sophisticated calculations in the framework of a model with $B_{ef} = 0$ [4] give a weaker dependence $\langle w^2(\mathbf{r}) \rangle \sim D^{2\varsigma}$ with $\varsigma \approx 0.6$, which, however, does not save the situation at $D \to \infty$.

In Fig. 2 we show the temperature dependencies of $\sqrt{\langle w^2(\mathbf{r})\rangle}$ calculated by formula (20) for graphene sheets of different sizes with $\kappa = 2.0$ eV and temperature-dependent graphene density and elastic parameters (curves 1–6); the results of numerical simulations are designated by empty symbols [16] and by full symbols [30]. Taking into account that Eq. (20) was derived in macroscopic limit it is reasonable to expect only qualitative agreement between our theory and numerical experiments for graphene sheets of small sizes (full symbols in Fig. 2), for which the boundary effects can be essential. However, it is possible to achieve reasonable quantitative agreement for curves (1–4) multiplying the nominal graphene size by a factor of ≈1.5. At the same time there is no need to introduce such fitting multiplier for graphene samples with more that ≈3000 atoms (empty symbols [16] on curve (5), 5000 atoms, and on curve (6), 33600 atoms). It would be interesting to test the theoretical predictions experimentally on quasi-macroscopic samples in wide range of temperatures (see [19, 20]).

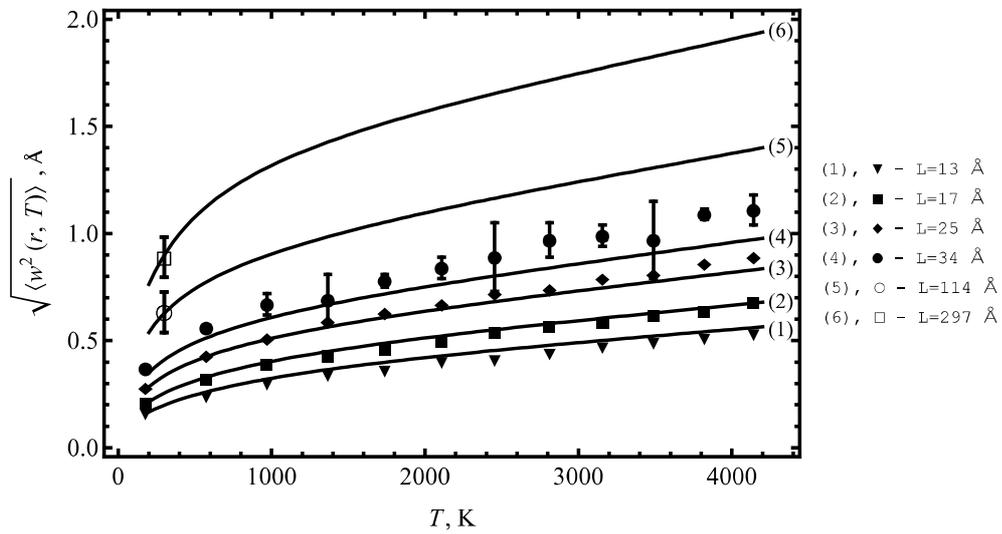

Fig. 2. Temperature dependences of $\sqrt{\langle w^2(\mathbf{r})\rangle}$ calculated by formula (20) for graphene sheets of different sizes $L$ (curves 1–6; the details of calculations are given in text); full and empty symbols denote the results of computer simulations of $\sqrt{\langle w^2(\mathbf{r})\rangle}$ from [30] and [16], respectively.

Another important characteristic – the average wavelength $\langle \lambda_w \rangle$ of the intrinsic ripples (microscopic corrugations of a graphene sheet) – is a quantity whose dependence on the graphene thermodynamic state can not so far be extracted even from the numerical "experiments". Following [21], define the average wavelength of the corrugations as

$$\langle \lambda_w \rangle \equiv 2\pi\sqrt{\langle w^2(\mathbf{r})\rangle/\langle [\nabla w(\mathbf{r})]^2\rangle}. \qquad (21)$$

It is important to note that just the values $\sqrt{\langle w^2(\mathbf{r})\rangle}$ and $\langle \lambda_w \rangle$, as the basic characteristics of the so-called corrugations of the "plane" graphene surface, were detected by the transmission electron microscopy technique on the free-standing graphene [21] and found in the numerical simulations of the 2D crystal dynamics [7, 11].

---

[4] When passing from $L$ to our $D$ we used the equality of graphene sheet areas: $L^2 \to \pi D^2/4$.

In Fig. 3 the temperature dependences of $\langle \lambda_w \rangle$, calculated using expressions (19) and (20), for different graphene sizes are shown. A distinctive feature of the calculated temperature dependences of $\langle \lambda_w \rangle$ is a maximum at $T \approx 500$ K, and it would be interesting to test this theoretical prediction in real experiments on macroscopic graphene samples. The characteristic values of $\langle \lambda_w \rangle \approx 50-100$ Å in Fig. 3 are in satisfactory agreement with those measured in the electron microscopy experiments on the free-standing graphene of large sizes [19, 20, 21] (for example, the value $\langle \lambda_w \rangle \sim 100$ Å have been found in [21]) and obtained in computer simulation of the dynamics of quasi-macroscopic fragments of the graphene crystal [7, 30–32].

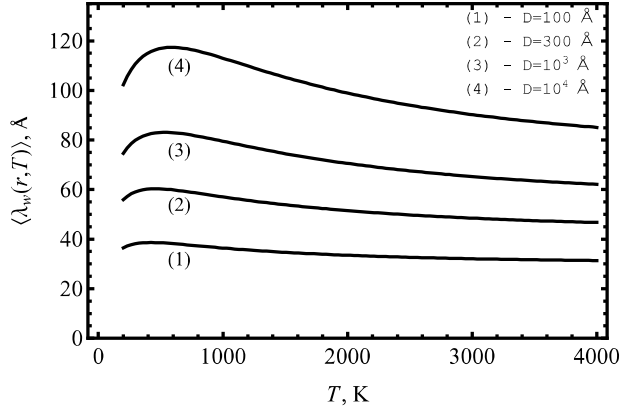

Fig. 3. Characteristic wavelength $\langle \lambda_w \rangle$ of microscopic corrugations as a function of temperature calculated by Eq. (21) using expressions (19) and (20) for different diameters of graphene sheet.

## 4. Conclusions

On the basis of transparent theoretical arguments, it is shown that for small wave numbers the dispersion of bending mode in 2D graphene-type crystals is linear, i.e. is "sound-like". This conclusion is consistent with the results of [12, 13], where the existence of the sound segment of the bending mode was established earlier from other positions. In the approach developed here the decisive role in the formation of bending sound in graphene play the terms of form (4) in the graphene strain tensor. The speed of the "bending sound" $s_B$ is determined by the elastic modulus $B$, which is calculated by averaging those terms in the "Hamiltonian" of elastic graphene strains over the thermal fluctuations of the "fast" in-plane acoustical oscillations. The obtained in this way $s_B$ is at least 15–20 times less than the in-plane mode speeds; with growing temperature the bending rigidity "toughens". Our results are in a rather good agreement with results of MD simulations [16] obtained recently for graphene sheets containing ≈30000 atoms. As can be concluded, based on the results [12], the account of anharmonic terms (3) and (12), which we ignored, may slightly modify $s_B$ at $T \to 0$ whereas at high temperatures $s_B$ is determined by quartic term (5) only. At the same time the account of contributions (3) and (12) considered in [12, 13] allows to renormalize the value of bending rigidity at low and high temperatures. The "sound" segment in the bending mode spectrum provides only logarithmic mean-square fluctuations of ripples amplitudes, like those of in-plane displacements, what is a common trait of 2D crystals. The proposed approach, in fact, does not require any parameters except of the density, initial bending rigidity $\kappa$, and in-plane elastic constants, known from independent sources. With the help of the obtained results, it is possible to express the average amplitude and wave-length of the microscopic corrugations of a graphene sheet through those parameters and temperature. The comparison of the theory with the data of real experiments and computer simulations demonstrates their good quantitative agreement in wide temperature interval and different sizes of graphene sheets for physically reasonable values of graphene in-plane elastic constants and $\kappa$. The developed approach, in principle, can be extended to other graphene-like crystals (such as silicene etc., [22–24]), but lack of knowledge on their mechanical parameters does not allow yet to do this.

**Acknowledgements**

This work was supported by the Ministry of Education and Science of Ukraine, Grants #0115U003208 and #0115U003214.

Authors are grateful to an anonymous reviewer for comments and questions, which allowed to improve the performance of the paper and to deepen to some extent its content. We are also grateful to the authors of paper [16] for the opportunity to know their results prior to publication.